\documentstyle[aaspp4]{article}

\def\stacksymbols #1#2#3#4{\def\theguybelow{#2}
	\def\verticalposition{\lower#3pt}
	\def\spacingwithinsymbol{\baselineskip0pt\lineskip#4pt}
	\mathrel{\mathpalette\intermediary#1}}
\def\intermediary #1#2{\verticalposition\vbox{\spacingwithinsymbol
	\everycr={}\tabskip0pt
	\halign{$\mathsurround0pt#1\hfil##\hfil$\crcr#2\crcr
		\theguybelow\crcr}}}
\def\lta{\stacksymbols{<}{\sim}{2.5}{.2}}
\def\gta{\stacksymbols{>}{\sim}{3}{.5}}

\begin{document}
\title{X-RAY OBSERVATIONS AND THE STRUCTURE OF 
ELLIPTICAL GALAXIES$^1$}

\author{Fabrizio Brighenti$^{2,3}$ and William G. Mathews$^3$}

\affil{$^2$Dipartimento di Astronomia,
Universit\`a di Bologna,
via Zamboni 33,
Bologna 40126, Italy\\
brighenti@astbo3.bo.astro.it}

\affil{$^3$University of California Observatories/Lick Observatory,
Board of Studies in Astronomy and Astrophysics,
University of California, Santa Cruz, CA 95064\\
mathews@lick.ucsc.edu}






\vskip .2in

\begin{abstract}
We compare optical and high quality 
x-ray data for three bright ellipticals 
in the Virgo cluster, NGC 4472, 4649, and 4636.
The distribution of 
total mass in NGC 4472 and 4649 
determined from x-ray data is sensitive
to the stellar mass over a considerable range in galactic radius 
extending to $r \approx r_e$, the effective radius.
The agreement of x-ray and optically-determined stellar masses 
provides a unique verification of the stellar 
mass to light ratio which is essentially constant 
over the range $0.1 \lta r/r_e \lta 1$.
However, for NGC 4636 the 
dark matter is important at all radii $\gta 0.35r_e$.
Evidently the dark to stellar mass ratio 
varies in quite different ways in ellipticals
of comparable optical luminosity, implying that 
the radial structure of dark halos may not be universal.
There is some evidence in NGC 4636 for additional 
support of the hot interstellar gas at $r \lta 0.35 r_e$;
either a field
$B \sim 10^{-4}$ G or 
a small (mechanically unstable) 
central region of high gas temperature 
($T \sim 10^7$ K) is required.
The global temperature structure in the hot 
interstellar medium of many recently observed ellipticals 
is very similar, reaching a maximum near $3 - 4 r_e$.
This feature, which may suggest a new structural scale in these 
galaxies, is inconsistent with current 
theoretical gas dynamical models.
\end{abstract}


\section{INTRODUCTION}

Elliptical galaxies are large, bright stellar systems with
considerable structural regularity. The stellar light profiles are 
well fit by the de Vaucouleurs law (Burkert 1993) 
and the stellar velocity dispersion, surface brightness, and effective 
radius $r_e$ are constrained to a fundamental plane (Djorgovski \& Davis
1987; Dressler et al. 1987).
By contrast, the x-ray emission from hot interstellar gas in 
ellipticals exhibits enormous 
variations in $L_x/L_B$ (Eskridge et al 1995) and, to a lesser 
extent, in gas temperature (Davis \& White 1996).

We have re-examined optical and x-ray observations of
ellipticals in preparation 
for a new series of gas dynamical studies of 
the evolution of the interstellar gas.
Although current x-ray observations are sparse, we have found that
they reveal more useful information about elliptical galaxies
than is currently realized.
Gas temperature profiles $T(R)$ for
elliptical galaxies recently determined with ROSAT and ASCA 
reveal a surprising uniformity for radii $r \lta 10 r_e$.
The total mass and density of all gravitating matter,
$M_{tot}(r)$ and $\rho_{tot}(r)$, can be found 
from the variation of temperature and density in the hot 
interstellar gas by assuming that the 
gas is in hydrostatic equilibrium in the galactic
potential.

We show here that 
the total mass determined from the x-ray gas 
for two bright Virgo ellipticals (NGC 4472 and 4649) is 
in excellent agreement with a de Vaucouleurs profile
for $0.1r_e \lta r \lta r_e$
using optically determined mass to light ratios.
Remarkably, in another bright Virgo elliptical NGC 4636 
dark matter contributes substantially to 
the total mass within $r_e$ 
and the total mass found from x-ray observations
is {\it lower} than the known stellar mass for
$r \lta 0.35 r_e$, implying that thermal pressure is 
not the only support for the hot gas.
These results suggest 
that AXAF x-ray observations of elliptical
galaxies will provide powerful new constraints 
on both the interstellar physics and the nature of 
stellar populations in these galaxies.

\section{SCALE OF THERMAL STRUCTURE IN ELLIPTICALS}

Figure 1 shows a plot of the radial variation of gas temperature with 
projected galactic radius 
$R/r_e$ for six early type galaxies.
The remarkable feature that all galaxies share is a positive
temperature 
gradient out to about $\sim 3r_e$ followed by a leveling 
off or gradual decrease toward larger radii.
Data for Figure 1 are taken from the following sources:
NGC 1399: Jones et al. (1997); 
NGC 5044: David et al. (1994);
NGC 4636: Trinchieri et al. (1994); 
NGC 507: Kim \& Fabbiano (1995);
NGC 4472: Irwin \& Sarazin (1996);
NGC 4649: Trinchieri et al. (1997).
The temperature profile observed
with the ASCA satellite by Mushotsky et al. (1994) 
for NGC 4636 is
in excellent agreement with that of Trinchieri et al. (1994)
for 20 percent solar abundance.
Additional properties of these galaxies are listed 
in Table 1.
It is remarkable that $T(R)$ 
is so consistent
in spite of the wide range of luminosities and
cosmic environments among these galaxies.
NGC 4636, 4649 and 4472 are in the Virgo cluster 
while NGC 1399, 5044, and
507 are the brightest galaxies in small groups or clusters.
NGC 507 has an exponential surface brightness profile that 
is consistent with a very massive, face-on S0
galaxy (Magrelli, et al. 1992).

The temperature maximum 
observed in 
ellipticals at $\sim 3r_e$ suggests a new previously unrecognized 
structural scale length.
This cooling of gas within $3 - 4 r_e$ is {\it not} a 
natural result of galactic cooling flows as many have suggested.
In theoretical models,
the radial gas velocity in the 
hot interstellar gas is very subsonic since otherwise
the gas replenishment time would be impossibly short and $L_x$
would be much too low.
Straightforward spherical models of the hot interstellar gas 
(e.g. Fig. 3 of Brighenti \& Mathews 1996) 
typically result in subsonic inflow with 
$dT/dr < 0$ for $r \gta 0.1 r_e$. 
Although the gas loses thermal energy by emitting the x-rays 
observed, it is heated by compression 
as it descends deeper into the galactic potential,
i. e. $dT/dr < 0$. 
In this sense the term ``galactic cooling flow'' is a misnomer.
In ``mass drop-out'' models, in which gas is removed from the inner 
flow, the remaining gas is actually heated since the $Pdv$ work 
done by gravity must raise its temperature even higher to provide 
hydrostatic support for gas at larger radii.
However, this rise in $T(r)$ is fully compesated when 
radiation from the cooler, dropping-out gas is included
(e.g. White \& Sarazin 1987).
In any case, mass dropout is usually invoked at small galactic
radii, less than the scale of $dT(r)/dr < 0$ in Figure 1.
The gas temperature is also 
sensitive to heating by supernova explosions, 
but this heating should be proportional to the stellar 
density so $dT/dr < 0$ at $r \lta r_e$ would again be expected.
To our knowledge no theoretical cooling flow model can produce 
temperature profiles resembling those in Figure 1.
Fortunately, regardless of the possibly complex thermal
history of the interstellar gas, 
the observed variation of $T$ and $\rho$ 
can be used to determine the underlying 
mass distribution as we now discuss.

\section{COMPARISON OF X-RAY AND OPTICAL DATA FOR THREE VIRGO GALAXIES}

NGC 4472, 4649 and 4636 listed in Table 1 are all members of the Virgo
cluster.
We adopt a distance of $D = 17$ Mpc for all three galaxies, 
similar to the 
Malmquist-corrected value $D = 17.2 \pm 1.9$ Mpc 
of Gonzalez \& Faber (1997).
For any particular Virgo galaxy the distance is uncertain by
$\sim 20$ percent due to the unknown location of the galaxy along
the line of sight through the cluster.

The total mass $M_{tot}(r)$ in these 
galaxies can be found from 
the condition for hydrostatic equilibrium:
$$M_{tot}(r) = - {k T(r) r \over G \mu m_p } 
\left( {d \log \rho \over d \log r} + {d \log T \over d \log r}
+ \phi_m {d \log P_m \over d \log r} \right)$$
where the last term allows for the possibility of 
magnetic pressure $P_m = B^2/8 \pi$ and $\phi_m =
P_m/P$ is the ratio of magnetic to gas pressure.
The proton mass is $m_p$ and we assume $\mu = 0.63$.
The strong negative density gradient $d \log \rho / d \log r$ 
is expected to be the largest of the three 
derivatives.
Only the shape of $\rho(r)$, not its 
absolute normalization, influences $M_{tot}(r)$.
The density distribution $\rho(r)$ in the hot interstellar gas
can be found from the x-ray surface brightness
distribution.
$T(r)$ and possibly also $\phi_m(r)$
must be known to some precision in order to 
determine reliable total masses.
The total mass density is 
$\rho_{tot}(r) = d M_{tot} / 4 \pi r^2 dr.$

Density profiles are available for all three ellipticals 
from {\it Einstein}
HRI observations (Trinchieri, Fabbiano \& Canizares 1986) and 
from ROSAT PSPC (Trinchieri et al. 1994; Irwin \& Sarazin 1996; 
Trinchieri, Fabbiano, \& Kim 1997). 
Fortunately there is a substantial range of angular scale 
over which these data 
sets overlap so values of $\rho(r)$ from {\it Einstein} and 
ROSAT data can be renormalized to agree.
Having done this we fit $n(r) = \rho(r)/m_p$ with a sum of functions 
$n(r) = \Sigma n_i(r)$ where $n_i(r) = n_{o}(i) [1 +
(r/r_{o}(i))^{p(i)}]^{-1}$ and the temperature is fit with 
$T(r) = 2 T_m [ r_m/(r + r_{ot}) + (r/r_m)^q]^{-1}$.
The temperature $T(R)$ 
observed at any projected radius $R$ is an average
along the line of sight weighted by $\rho^2$
and differs in principle from 
the temperature $T(r)$ at physical radius $r$. 
However, $\rho^2$ is a very steep function of
galactic radius and we find $T(r) \approx T(R)$ within 
10 percent, sufficient for our purposes here considering
the observational uncertainties involved (Fig. 1).
Figures 2a, 2d and 2g show our fits to the x-ray data. 
(The parameters for these fits are:
NGC 4472: $n_o(i) =$ 0.095, 0.00597, -0.0004;
$r_o(i) =$ 0.107, 0.95, 10.; $p(i) =$ 2.0, 1.14, 1.19;
$T_m =$ 0.75; $r_m =$ 0.5; $r_{ot} =$ 0.75; $q =$ 0;
NGC 4636: $n_o =$ 0.151; $r_o =$ 0.172; $p$ = 1.57;
$T_m =$ 0.5375;
$r_m =$ 0.475; $r_{ot} =$ 0.6; $q =$ 0.0;
NGC 4649: $n_o(i) =$ .1, .0014; $r_o(i) =$ .15, 3.0;
$p(i) =$ 1.8, 3.0;
$T_m =$ 0.9; $r_m =$ 4.0; $r_{ot} =$ 4.0; $q =$ 0.0
with radii in $r_e$, densities in cm$^{-3}$ and
temperatures in $10^7$ K.)
Total masses $M_{tot}(r)$ 
(Figs. 2b, 2e and 2h) and the corresponding 
total mass densities $\rho_{tot}(r)$ 
(Figs. 2c, 2f, and 2i)
are determined with $\phi_m(r) = 0$.
Our values for $M_{tot}(r)$ in Figure 2b 
are in satisfactory agreement with 
those of Irwin and Sarazin (1996) (scaled to 
$D = 17$ Mpc) which are 
based only on ROSAT data; our $M_{tot}(r)$ in Figure 2h
is in excellent agreement with 
Mushotsky et al. (1994).

It is of particular interest to compare $M_{tot}(r)$ and 
$\rho_{tot}(r)$ with corresponding stellar values.
The total mass $M_{*t}$ 
is found from $L_B$ using ``stellar'' 
mass to light ratios of van der Marel (1991), determined by 
comparing stellar velocities
measured out to $r_{obs} \approx 0.5 r_e$ with solutions of 
the 2D Jeans equations in slowly rotating 
model galaxies.
This ``stellar'' $M_{*t}/L_B$ 
is sensitive to {\it all} mass within
$r_{obs}$, and may contain a component of dark matter which 
can be checked {\it a posteriori} (see below).
In general 
mass to light values determined only in the core region
(e.g. Faber et al. 1997) are larger; 
central black holes may account for this increase 
but typical holes may not be sufficiently 
massive (Kormendy, J. \& Richstone, D. 1995;
Faber et al. 1997).
$M_*(r)$ and $\rho_*(r)$ in Figure 2 
are evaluated with 
de Vaucouleurs profiles (Young 1976).

{\it The Structure of NGC 4472 and 4649}:
We find it quite remarkable that the total 
mass and density indicated by 
the x-ray observations agree almost exactly with 
stellar values in the range $0.1 \lta r \lta 1r_e$
where the observational data is of high quality
(Figs 2b and 2e).
Discounting a conspiracy of compensating errors, 
this superb agreement is possible 
only 
(i) if the stellar mass to light values determined by 
van der Marel (1991) 
are essentially correct and constant for stars out 
to $\sim 1 r_e$, 
and (ii) if 
magnetic pressure or rotation 
contributes little or nothing to the total
pressure support of gas in $0.1 \lta r \lta 1r_e$.
The total masses of dark and stellar matter are equal at 
$\sim 2.5 r_e$ which includes $r_{obs}$ so 
van der Marel's $ M_*/L_B$ are likely to be totally stellar. 
Dark matter dominates at $r/r_e \gta 2.5$.
At $r = 10r_e$ the total mass to light ratio is $\sim 78$ 
(NGC 4472) and $\sim 110$ (NGC 4649)
and $M_{tot}/L_B$ is undoubtedly higher at $r/r_e \gta 10$.
The stellar to dark halo 
transition is particularly striking in the density plots
(Figs 2c and 2f).
None of the qualitative features in Figure 2 are changed if
$(M_*/L_B)_{core}$ values are used 
from Faber et al. (1997), but the overall fit in 
$0.1 \lta r/r_e \lta 1$ is of lower quality. 
The discrepancy $M_{tot} < M_*$ at
$r \lta 0.1 r_e$ may result from observational
inaccuracies in this region 
although $M_{tot}(r)$ 
would be underestimated if a large magnetic field were present. 

{\it The Structure of NGC 4636}: 
Our total mass $M_{tot}(r)$ for NGC 4636 (Fig. 2h) 
is in excellent agreement with the mass determined
by Mushotsky et al (1994) who also assumed $D = 17$ Mpc. 
A de Vaucouleurs profile is used to determine $M_*(r)$ 
and $\rho_*(r)$;
we have verified that 
NGC 4636 satisfies a de 
Vaucouleurs profile by plotting the surface brightness data
of Peletier et al (1990) against $R^{1/4}$.

The most surprising result evident from Figures 
2h and 2i is the relative 
dominance of dark matter in NGC 4636 as compared to NGC 4472
and 4649.
The dark mass $M_{dark} \equiv M_{tot} - M_*$ 
in NGC 4636 becomes equal to the stellar mass $M_*$
at about $r = 1.2 r_e$ and $M_{dark} \approx 0.83 M_*$ 
at $r = r_e$ where the total mass to light ratio 
is 19.
This large amount of dark matter may be unusual since it has 
been difficult until recently 
to find evidence of dark matter in ellipticals from stellar 
velocities measured within $r_e$ 
(e.g. Carollo et al. 1995).
The ``stellar'' mass to light ratio $M_*/L_B$ from 
van der Marel is
determined from velocity observations only out to
$r_{obs}/r_e = 0.45$, where $M_* \gg M_{dark}$ (Fig. 2h).
It is remarkable that there is no change in slope in Figure 2h 
as $M_{tot}(r)$ crosses below $M_*(r)$ 
at $r \approx 0.33 r_e$ since 
all observations should be reliable at this radius.
This failure of the x-ray determined mass 
to detect the stellar mass suggests large magnetic 
fields, unusually high gas temperatures 
or rotational support in this region.
Magnetic fields $B \sim 9 \times 10^{-5}$ G would be required
at $r = 0.1r_e$ to support the hot gas against the 
stellar mass within (see Mathews \& Brighenti 1997).
Or a curious central temperature inversion may exist: 
a mean temperature within 1 kpc (0.20') of 
$0.9 \times 10^7$ K is implied.
This inversion is constrained so that
the mean projected gas temperature within 1',
$0.75 \times 10^7$, is within the observed limits 
$0.667 +0.029/-0.041 \times 10^7$ K.
However, the high temperature interpretation 
for $M_{tot} < M_*$ in NGC 4636 is unlikely 
since it is buoyantly unstable; $dT/dr$ is superadiabatic 
for $r \lta 0.08 r_e \approx 0.7$ kpc.
Alternatively, $M_*/L_B$ could decrease with 
galactic radius so that $M_*$ is lower than we think near 
$r/r_e \approx 0.1$ in Figure 2h, 
but this seems unlikely in view of the 
uniform $M_*/L_B$ implied by Figures 2b and 2e.
None of these conclusions are changed if the 
core value $(M_*/L_B)_{core} = 12.69$ is used instead.
The different distribution of dark matter at $r \lta r_e$ 
in NGC 4636 may provide evidence against a universal 
dark halo structure as proposed by 
Navarro, Frenk \& White (1996), but for $r \gta r_e$ the 
dark halos are all very similar having slope 
$\rho_{dark} \propto r^{-1.9}$ at $r \sim 10r_e$.
The total mass to light ratio near the outer limit of
the x-ray observations shown in Figure 2g,
$r = 12.5r_e$, is about 125.

How do these Virgo ellipticals differ in other respects?
NGC 4636 is a relatively isolated galaxy quite far 
($\gta 3$ Mpc) from the core of Virgo.
Nevertheless, its x-ray image shows some
azimuthal asymmetry at $r \gta 5r_e$, but this emission 
may arise from a different source (Trinchieri et al. 1994).
It also has a broad, faint stellar distribution characteristic 
of CD galaxies.
NGC 4472 is optically the brightest galaxy in the Virgo cluster,
lies within a small subgroup, and is interacting with a 
nearby dwarf irregular galaxy, UGC 7636.
At radii $\gta 3r_e$ the x-ray isophotes show a tail-like structure
which may result from the motion of NGC 4772 through the Virgo
cluster medium (Forman et al 1985; 
Irwin and Sarazin 1996).
NGC 4472 has a small kinematically decoupled (non-rotating) core
(Davies 1989) within $\sim 0.096r_e$.
NGC 4649 appears to be close to the center of Virgo.
However, none of these galaxies shows evidence of unusual 
x-ray or optical structure at $0.1 \lta r/r_e \lta 3$ where the 
difference in dark to stellar mass is so apparent in Figure 2.

\section{CONCLUSIONS}

(1) A previously unrecognized spatial scale at $\sim 3r_e$ is
suggested by the similarity of gas temperature profiles in 
many recently observed ellipticals.
(2) The stellar mass to light ratios for NGC 4472 and 4649  
have been verified by x-ray observations; the interstellar 
gas temperature is also correct.
(3) The mass to light ratio in NGC 4472 and 4649 determined 
from stellar motions are constant over the 
range $0.1r_e \lta r \lta r_e$.
(4) In NGC 4472 and 4649 the hot interstellar gas is supported out to 
$r \sim r_e$ by thermal gas pressure, other means of support 
such as magnetic pressure or rotation are not evident.
(5) Hot interstellar gas in the center of 
NGC 4636 may be supported by magnetic stresses 
($B \sim 10^{-4}$ G), a transient unstable region of 
unexpectedly high temperatures ($T \sim 10^7$ K), or rotation.
(6) Elliptical galaxies with comparable luminosities 
can have quite different distributions of
dark matter relative to stellar matter. 
(7) In NGC 4636 the mass of dark matter is comparable 
to that in stars at $r \sim r_e$; in NGC 4472 and 4649 the mass of 
dark halo matter is negligible at $r \sim r_e$.
(8) The ``stellar'' mass to light ratios for these three 
ellipticals as determined by van der Marel (1991) 
are unlikely to be contaminated by dark matter from the 
galactic halos. AXAF observations will be useful
in interpreting the variation of $M_*/L_B$ along the fundamental 
plane (Pahre \& Djorgovski 1997) and in verifying central 
magnetically supported regions.

\acknowledgments

Thanks to Michael Loewenstein and Sandra Faber for enlightenment 
and encouragement.
Our work on the evolution of hot gas in ellipticals is supported by
NASA grant NAG 5-3060 for which we are very grateful. In addition
WGM is supported by a UCSC Faculty Research Grant and FB is supported 
in part by Grant ASI-95-RS-152 from the Agenzia Spaziale Italiana.



\figcaption[aasxobsfig1.ps]{Radial dependence of gas temperature
with projected radius in five early
type galaxies all showing maxima at $\sim 3r_e$. \label{fig1}}

\figcaption[aasxobsfig2.ps]{ (a,d,g) Observations and our fits 
to gas temperature (top)
and density (bottom) for {\it Einstein} HRI (filled circles) and
ROSAT (open circles) data;
(b,e,h) {\it solid curves}: $M_{tot}$; points are values of $M_{tot}$
from Irwin and Sarazin (1996) for 4472 and Mushotsky et al. (1994) for
4636;
{\it long-dashed curves}: stellar mass $M_*$;
{\it short-dashed curves}: dark mass
$M_{dark} \equiv M_{tot} - M_*$ which
loses accuracy when $M_* \ge M_{tot}$;
{\it dot-dashed curves}: total mass of hot gas;
(c,f,i) {\it solid curves}: total mass density $\rho_{tot}$;
{\it long-dashed curves}: stellar mass density $\rho_*$ with a
horizontal cut at the core or break radius $r_b$;
{\it short-dashed curves}: dark mass density. \label{fig2}}


\makeatletter
\def\jnl@aj{AJ}
\ifx\revtex@jnl\jnl@aj\let\tablebreak=\nl\fi
\makeatother

\begin{deluxetable}{rcrrrlrll}
\tablewidth{0pt}
\tablecaption{OPTICAL AND X-RAY PROPERTIES OF SIX ELLIPTICALS}
\tablehead{
\colhead{NGC} & 
\colhead{Type\tablenotemark{a}} &
\colhead{$D$\tablenotemark{b}} & 
\colhead{$L_B$\tablenotemark{c}} & 
\colhead{$M_{*t}/L_B$\tablenotemark{c,d}} &
\colhead{$L_x$\tablenotemark{c}} & 
\colhead{$r_e$\tablenotemark{c}} & 
\colhead{$r_e$} & 
\colhead{$r_b$\tablenotemark{e}} \nl
\colhead{} &
\colhead{} &
\colhead{(Mpc)} &
\colhead{($10^{10}$ $L_{B\odot}$)} &
\colhead{} &
\colhead{($10^{41}$ e/s)} &
\colhead{(kpc)} &
\colhead{(')} &
\colhead{('')} 
}
\startdata
507 & SO & 72.2 & 19.81 & ... & 59.3 & 26.94 & 1.285 & ... \cr
1399 & E1 & 17.9 & 2.80 & 10.85 & 8.25 & 3.63 & 0.706 & 3.14 \cr
4472 & E2 & 17 & 7.89 & 9.20 & 4.54 & 8.57 & 1.733 & 2.41 \cr
4649 & E2 & 17 & 5.35 & 9.03 & 1.61 & 6.07 & 1.227 & 3.58 \cr
4636 & EO+ & 17 & 3.54 & 10.74 & 3.79 & 8.32 & 1.683 & 3.21 \cr
5044 & E0 & 37.3 & 7.11 & ... & 52.6 & 14.24 & 1.315 & ... \cr

\tablenotetext{a}{Type from RC3.}
\tablenotetext{b}{Distances NGC 507 \& 5044 from Faber et al. (1989); 
NGC 1399 from Faber et al. (1997);
NGC 4636 and 4472 from Gonzalez \& Faber (1997).}

\tablenotetext{c}{Scaled to the distance in the second column.}

\tablenotetext{d}{Mass to light ratios from van der Marel (1991).}

\tablenotetext{e}{Break radii into core region from Faber et
al.(1997).}

\enddata
\end{deluxetable}


\begin{references}
\reference{head1988} Brighenti, F. \& Mathews, W. G. 1996, ApJ, 470,
747
\reference{head1988} Burkert, A., 1993, A\&A, 278, 23
\reference{head1988} Carollo, C. M., de Zeeuw, P. T., van der Marel,
R. P.,
Danziger, I. J., \& Qian, E. E. 1995, ApJ, 441, L25
\reference{head1988} David, L. P., Jones, C., Forman, W., \& Daines,
S. 1994,
ApJ, 428, 544
\reference{head1988} Djorgovski, S. \& Davies, M. 1987, ApJ, 313, 59
\reference{head1988} Dressler, A., et al. 1987, ApJ, 313, 42
\reference{head1988} Eskridge, P. B., Fabbiano, G., \& Kim, D-W. 1995,
ApJ, 441, 182
\reference{head1988} Faber, S. M., Tremaine, S., Ajhar, E. A., Byun,
Y-I.,
Dressler, A, Gebhardt, K., Grillmair, C., Kormendy, J.,
Lauer, T. R., \& Richstone, D. 1997, ApJ (in press)
\reference{head1988} Faber, S. M., Wegner, G., Burstein, D., Davies,
R. L.,
Dressler, A., Lynden-Bell, D., \& Terlevich, R. J. 1989,
ApJS, 69, 763
\reference{head1988} Forman, W., Jones, C., \& Tucker, W. 1985, ApJ,
293, 102
\reference{head1988} Grillmair, C., \& Tremaine, S. 1996, AJ, 112, 105
\reference{head1988} Gonzalez, A. H. \& Faber, S. M. 1997, ApJ, (in
press)
\reference{head1988} Jones, C., Stern, C., Forman, W., Breen, J.,
David, L., \&
Tucker, W. 1997, ApJ, (in press)
\reference{head1988} Irwin, J. A. \& Sarazin, C. L. 1996, ApJ, 471,
683
\reference{head1988} Kim, D-W. \& Fabbiano, G. 1995, ApJ, 441, 182
\reference{head1988} Kormendy, J. \& Richstone, D. 1995, ARA\&A, 33,
581
\reference{head1988} Magrelli, G., Bettoni, D., \& Galletta, G.1992,
MNRAS, 256, 500
\reference{head1988} Mathews, W. G. \& Brighenti, F. 1997, ApJ, (in press).
\reference{head1988} Mushotzky, R., Loewenstein, M., Awaki, H,
Makishima, K,
Matsushita, K., \& Matsumoto, H. 1994, ApJ, 436, L79
\reference{head1988} Navarro, J. F., Frenk, C. S., \& White, S. D. M.
1996,
ApJ, 462, 563
\reference{head1988} Pahre, M. A. \& Djorgovski, S. G. 1997, in {\it
The Second
Stromlo Symposium: The Nature of Elliptical Galaxies},
eds. M. Arnaboldi, G. S. Da Costa, \& P. Saha,
ASP Conf. Ser. 116, 154
\reference{head1988} Peletier, R. F., Davies, R. L., Illingworth, G.
D., Davis, L. E., \& Cawson, M. 1990, AJ, 100, 1091
\reference{head1988} Trinchieri, G., Fabbiano, G., \& Canizares, C. R. 1996,
ApJ, 310, 637
\reference{head1988} Trinchieri, G., Fabbiano, G., \& Kim, D-W. 1997,
A\&A, 318, 361
\reference{head1988} Trinchieri, G., Kim, D-W., Fabbiano, G., \&
Canizares, C. R. C.
1994, ApJ, 428, 555
\reference{head1988} van der Marel, R. P. 1991, MNRAS, 253, 710
\reference{head1988} White, R. E. \& Sarazin, C. L. 1987, ApJ, 318,
621
\reference{head1988} Young, P. J. 1976, AJ, 81, 807
\end{references}
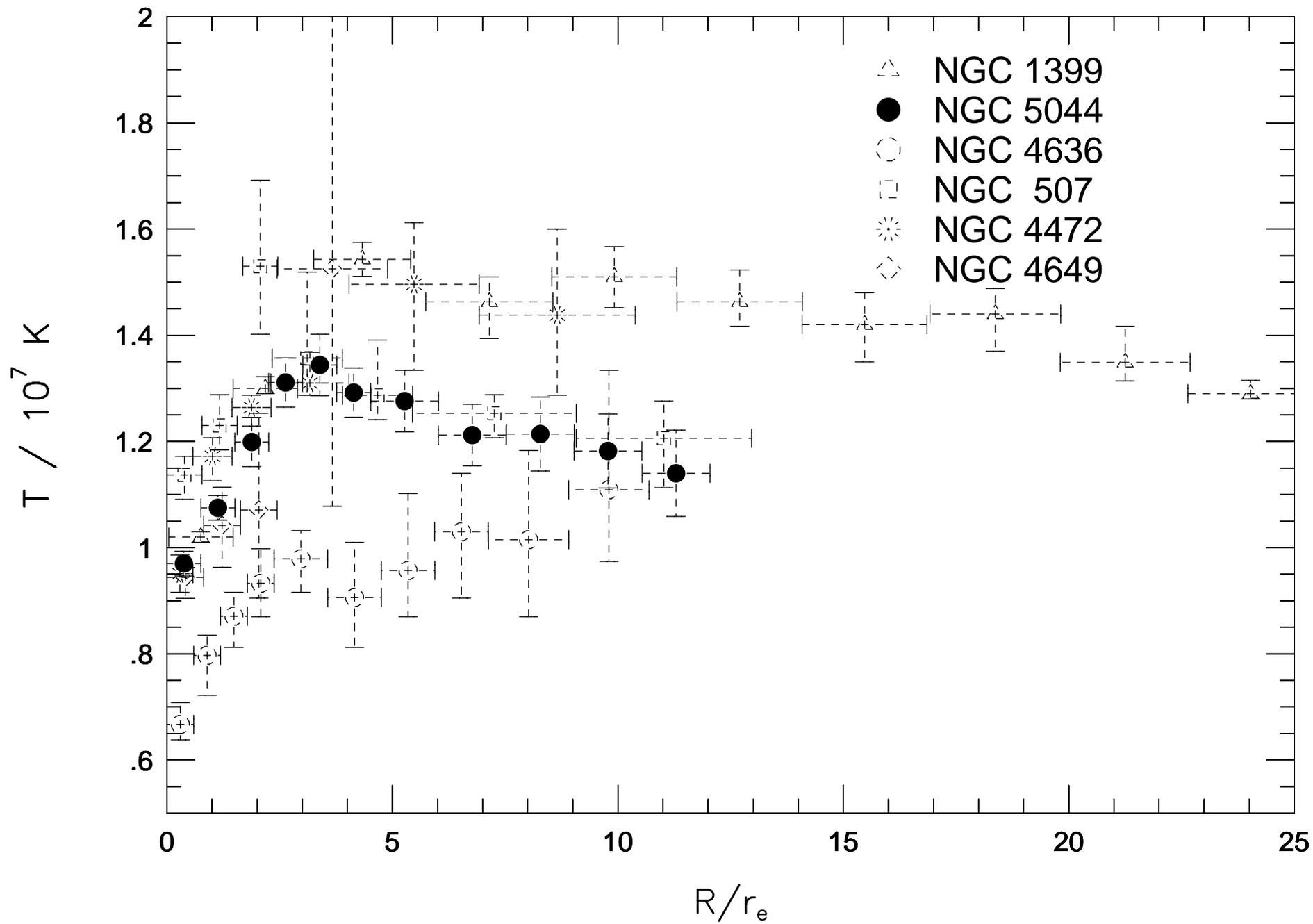
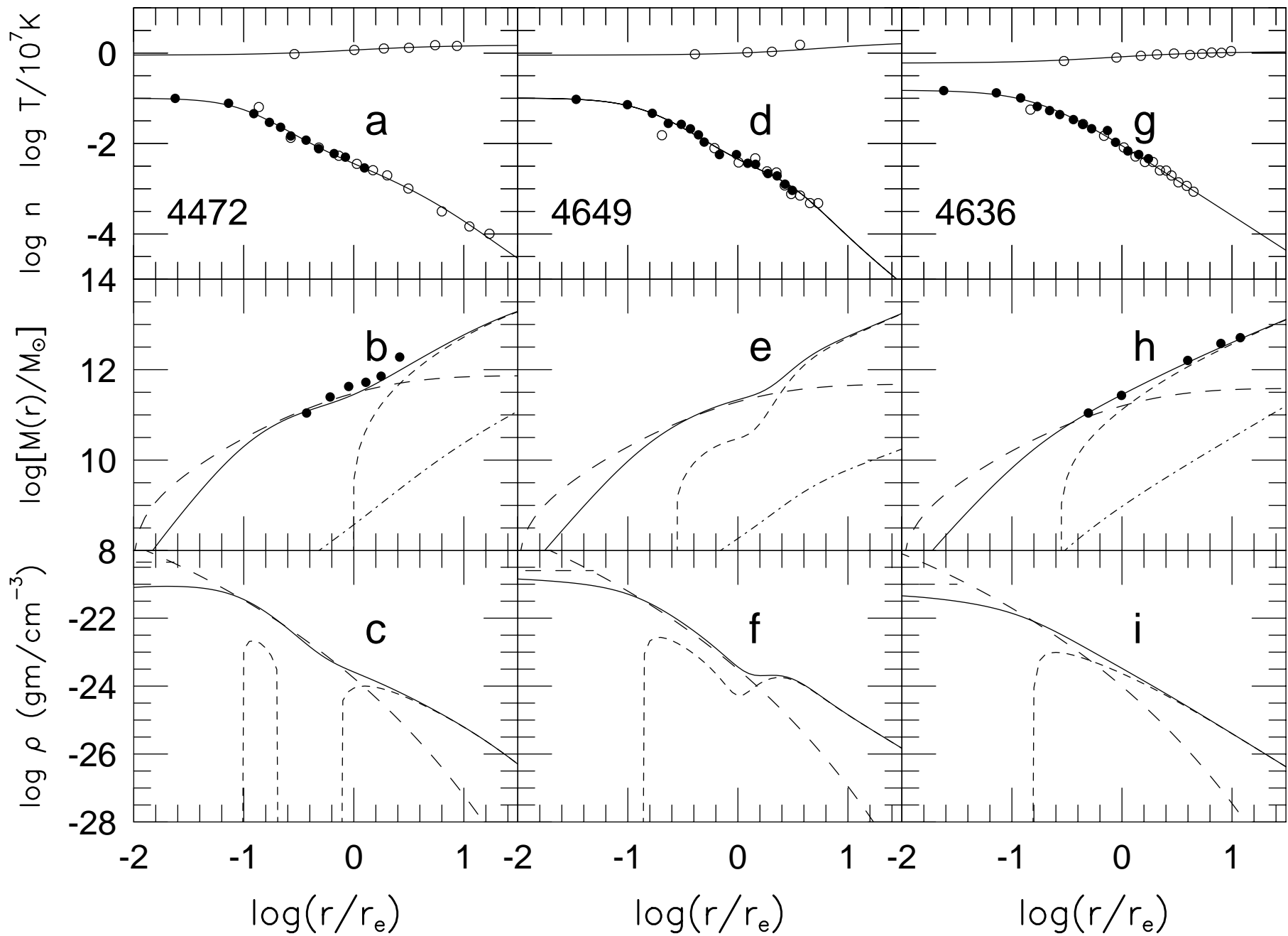
\end{document}